\documentclass[prl,superscriptaddress,twocolumn,floatfix,notitlepage,citeautoscript]{revtex4-2}
\usepackage[T2A]{fontenc}
\usepackage{amsmath}
\usepackage{graphicx}
\usepackage{color}
\usepackage{epstopdf}

\begin{document}
\title{Dark and thermal reservoir contributions to polariton sound velocity}

\author{Anna~M.~Grudinina}
\affiliation{National Research Nuclear University MEPhI (Moscow Engineering Physics Institute), 115409 Moscow, Russia}
\affiliation{Russian Quantum Center, Skolkovo IC, Bolshoy boulevard 30 bld. 1, 121205 Moscow, Russia}

\author{Nina~S.~Voronova}
\email{nsvoronova@mephi.ru}
\affiliation{National Research Nuclear University MEPhI (Moscow Engineering Physics Institute), 115409 Moscow, Russia}
\affiliation{Russian Quantum Center, Skolkovo IC, Bolshoy boulevard 30 bld. 1, 121205 Moscow, Russia}

\begin{abstract}
Exciton-polaritons in an optical microcavity can form a macroscopically coherent state despite being an inherently driven-dissipative system. In comparison with equilibrium bosonic fluids, polaritonic condensates possess multiple peculiarities that make them behave differently from well-known textbook examples. One such peculiarity is the presence of dark excitons which are created by the pump together with optically-active particles. They can considerably affect the spectrum of elementary excitations of the condensate and hence change its superfluid properties. Here, we theoretically analyze the influence of the bright and dark ``reservoir'' populations on the sound velocity $c_s$ of incoherently-driven polaritons. Both pulsed and continuous-wave pumping schemes characterized by essentially different condensate-to-reservoir ratio are considered. We show that the dark exciton contribution leads to considerable lowering of $c_s$ and to its deviation from the square-root-like behavior on the system's chemical potential (measurable condensate blueshift). Importantly, our model allows to unambiguously define the density of dark excitons in the system by experimentally tracking $c_s$ against the condensate blueshift and fitting the dependence at a given temperature.
\end{abstract}

\maketitle

Strong coupling between light and matter modes inside a semiconductor microcavity with embedded quantum wells results in the appearance of hybridized quasiparticles: exciton-polaritons. As interacting bosons, they demonstrate macroscopic coherence~\cite{RMP2010} and quantum properties~\cite{QFL}, and at the same time also the ability to propagate without friction, displaying the most remarkable collective phenomenon of superfluidity~\cite{amo_nature,amo,sanvitto_2010,sanvitto_superfluid}.

\begin{figure}[t]
\includegraphics[width=\columnwidth]{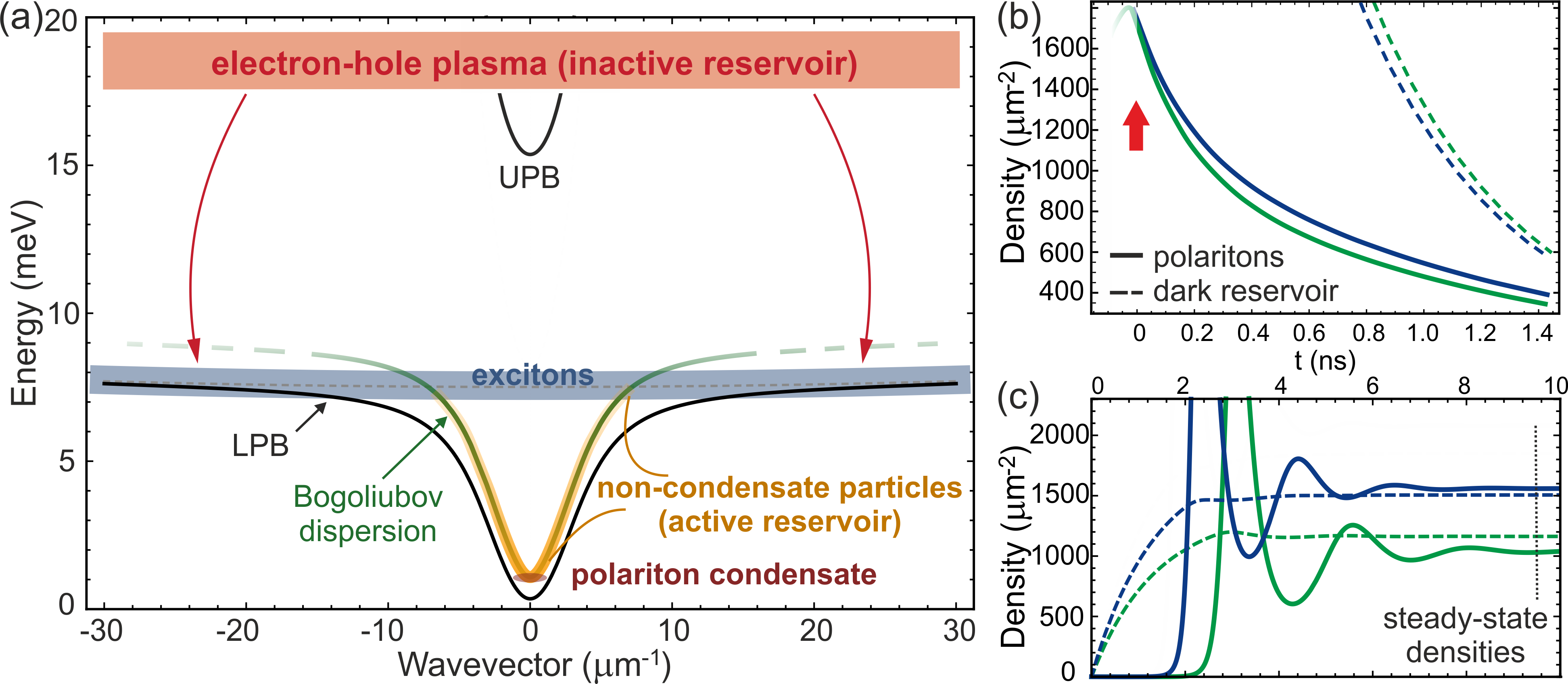}
\caption{\small (a) Schematic illustration of a non-resonantly pumped polariton system: hot electron-hole plasma (inactive reservoir) is relaxing towards both bright and dark exciton states, forming the populations of polaritons along the LPB (black solid line) and dark reservoir (gray shaded dashed line). Non-condensate polaritons in turn relax towards the lowest energy state to form a quasi-equilibrium condensate blueshifted due to interactions. The green solid line shows the Bogoliubov branch of excitations. (b,c) Temporal evolution of the polariton (solid lines) and dark reservoir (dashed lines) populations in the case of pulsed (b) and c.w. (c) pumping, according to the dynamical equations (\ref{nonres}). In (b), the red arrow indicates the moment of time when the densities reach the maximum after the arrival of the pulse. The initial polariton density is $n(0) = 1.8\times10^{11}$~cm$^{-2}$, the initial reservoir densities are $n_D(0)= 0.9$, $n_{in}(0)= 1.7\times 10^{12}$~cm$^{-2}$ (blue) and $n_D(0)= 1$, $n_{in}(0)= 1.5\times10^{12}$~cm$^{-2}$ (green); $R_{\rm p}^{-1}=3$~ns. In (c), $P=3P_{th}$ (green), $P=4P_{th}$ (blue), with $\tilde{P}=0.25P$ and $R_{\rm cw} = 0.1\times10^{-5}~\mu$m$^2/$ps. The other parameters for both cases $\gamma_{in}^{-1} = 2.5$~ns, $\gamma_D^{-1}=0.5$~ns, $D^{-1}=30$~ns, $\gamma^{-1} = 180$~ps.}
\label{fig1}
\end{figure}

Polariton Bose condensates are regularly achieved at elevated temperatures (from 4~K to room $T$) in a non-equilibrium setting \cite{kasprzak,balili,christmann,bloch2009}. Typically, an excitation laser is tuned either in resonance with the lower polariton branch (LPB) or high above the exciton dispersion, leading to formation of the so-called reservoir that relaxes to the LPB and eventually into the ground state. Due to finite lifetime, the polariton  population is decaying in time after the excitation pulse, or needs to be constantly replenished by means of continuous pumping.
Such driven-dissipative nature of the system leads to modifications of the elementary excitation spectrum
~\cite{wouters2007,yamamoto2012,haug2020,solnyshkov} and to violation of the Galilean invariance and Landau criterion~\cite{stepanov,amelio_prr,amelio_prb}, all of which are essential for understanding the specifics of polariton superfluidity.
The main challenge in building self-consistent descriptions is the impossibility to apply equilibrium microscopic approach developed by Bogoliubov~\cite{bogoliubov}. One is forced to either neglect the effects of gain and dissipation~\cite{grudinina} or build phenomenological descriptions based on the Gross-Pitaevskii-like equation (GPE) for the condensate coupled to the rate equation for the excitonic reservoir~
\cite{wouters2007,yamamoto2012,haug2020,solnyshkov,stepanov,amelio_prr,amelio_prb}. In the latter case, for a better correspondence with experimentally observed dynamics, the high-energy reservoir can be split into two parts~\cite{klagoud,veit,pieczarka17}, the `inactive' population directly created by the pump and the `active' reservoir that feeds the polariton condensate (the schematics is shown in Fig.~\ref{fig1}a). Ref.~\cite{anton} considers the model of three reservoirs: inactive, active, and dark. Dark, or spin-forbidden ($J=\pm2$), excitons are injected in the system at nonresonant excitation together with bright ($J=\pm1$) exciton populations (overall, there are four exciton branches per each quantum well filled from the relaxing charge carriers). Current studies show that such dark reservoir can be highly populated~\cite{menard} and long-lived~\cite{schmidt}. However its influence on polariton properties is mostly unexplored.

GPE-based approaches describe reasonably well the dynamics of macroscopically-populated polariton condensates. At the same time, the Bogoliubov spectra derived for the excitations on top of such driven-dissipative GPEs reveal either gapped or diffusive real part of the dispersion~\cite{wouters2007,yamamoto2012,haug2020}, which does not correspond to experimental observations to date~\cite{utsunomiya,kohnle,pieczarka2015,ballarini2020,pieczarka2020}. A combined GPE-Bolzmann equation for the condensate~\cite{solnyshkov} recovers the gapless, equilibrium-like linearization of the excitation spectrum at small momenta, with the sound velocity defined by the condensate density $n_0$: $c_s = \sqrt{gX^4_0n_0/m_{\rm LP}}$, where $m_{\rm LP}$ is the lower polariton effective mass, $g$ the exciton-exciton interaction constant and $X_{\bf p}$ the exciton Hopfield coefficient at momentum ${\bf p}$. A similar result was obtained in Refs.~\cite{stepanov,amelio_prr,amelio_prb,bramati_arxiv} for resonant excitation. In nonresonant pumping scheme experiments, however, the slope of this linear part was shown to depend on the chemical potential (observed condensate blueshift) controversially for different temperatures~\cite{kohnle,pieczarka2015}, and even to deviate from the square-root scaling law \cite{estrecho2021}. This controversy indicates that the effects of finite temperature and the optically-dark reservoir on the elementary excitations of polariton condensates are poorly understood.
The aim of the current Letter is to provide a stitching of the finite-temperature theory~\cite{grudinina} with a simple phenomenological rate-equations model that allows addressing the situations with large reservoir-to-condensate ratios~\cite{estrecho2021,parish}.

We start our theory with a brief discussion of the results derived in Ref.~\cite{grudinina} where non-condensed polaritons (active reservoir) as well as the reservoir of dark excitons are rigorously, rather than phenomenologically, included into the self-consistent Hartree-Fock-Bogoliubov (HFB) description of exciton-polariton gases at finite temperatures. It was shown that the non-parabolicity of the LPB, non-zero temperature, and the presence of dark excitons renormalize the Bogoliubov excitation spectrum. The corresponding sound velocity at ${\bf p}\to0$ is
\begin{equation}\label{cs_prb}
c_s = \sqrt{\frac{gX_0^4n_0}{m_{\rm LP}} \left(1+\frac{2\mu}{\sqrt{(\hbar\Omega)^2+\Delta^2}}\right)},
\end{equation}
where $\hbar\Omega$ and $\Delta$ are the Rabi splitting and the detuning of the cavity photon from the exciton resonance, respectively, and the chemical potential
\begin{equation}\label{mu_prb}
\mu = gX_0^2(X_0^2n_0+2n_X^\prime + n_D)
\end{equation}
contains contributions from the integrated dark exciton density $n_D$ and the density of non-condensed bright excitons at a given temperature
\begin{equation}\label{n'q}
n_X^\prime = \int X_{\bf p}^2\langle\hat{P}_{\bf p}^\dag\hat{P}_{\bf p}\rangle\frac{d{\bf p}}{(2\pi\hbar)^2}.
\end{equation}
The non-condensate part of the polariton field $\hat{P}_{\bf p}$ defining (\ref{n'q}) at the same time defines the integrated density of the active reservoir $n^\prime=\int\langle\hat{P}_{\bf p}^\dag\hat{P}_{\bf p}\rangle d{\bf p}/(2\pi\hbar)^2$ distributed along the whole non-parabolic LPB. This way, the active reservoir consists of non-condensate polaritons $n^\prime$, while only their excitonic fraction $n_X^\prime$ contributes to the blueshift. The density of polaritons is given by the sum $n=n_0+n^\prime$, and the total reservoir density is defined as $n_R = n^\prime + n_D$.
Within the model of Ref.~\cite{grudinina}, all the densities $n_0$, $n^\prime$, $n_D$ are equilibrium and dependent on the temperature.
The predictions (\ref{cs_prb}), (\ref{mu_prb}) and corresponding dependence $c_s(\mu)$ match well the experimental observations made at low $T$~\cite{kohnle,pieczarka2015}. At the same time, the assumption of equilibrium leads to $n_D\ll n$, and the theory fails to describe situations when different populations are defined by the dynamical equilibration of the system rather than by their thermal distributions.

Here, we consider the non-resonant excitation schemes and develop a unified description for cases when the dark exciton density is not small. As discussed in Ref.~\cite{amelio_prr}, the presence of a reservoir defines a privileged frame of reference, that associated with the static condensate. If one considers a moving polariton fluid, the effect of the reservoir on superfluid properties will be void as the condensate quickly leaves the area of space overlapping with the reservoir populations. For this reason, we limit ourselves to the most relevant case when the polariton fluid is at rest. 
Fig.~\ref{fig1}a shows schematically how the pump injects the high-energy population of charge carriers which cool down, filling available energy states on the exciton and lower-polariton dispersions. The inactive reservoir $n_{in}$ is therefore feeding both the dark exciton reservoir 
(all excitons not converting into polaritons) and the polariton subsystem on the exciton-like part of the LPB dispersion. We model the dynamics using the following equations:
\begin{eqnarray}
  \frac{\partial n_{in}}{\partial t} & = & P - \gamma_{in}n_{in} - Dn_{in} - R(n)n_{in}, \nonumber \\
  \frac{\partial n_D}{\partial t} & = & \tilde{P} - \gamma_Dn_D + Dn_{in}, \label{nonres} \\
  \frac{\partial n}{\partial t} & = & -\gamma n +R(n)n_{in},\nonumber
\end{eqnarray}
where $\gamma_{in}$, $\gamma_D$ and $\gamma$ are the decay rates of inactive reservoir, dark excitons, and polaritons ($\gamma\gg\gamma_{in}$), $D$ and $R(n)$ are the scattering rates from the inactive reservoir to dark excitons and to polaritons, respectively. Pump enters the equations for $n_{in}$ and $n_D$ as $P$ and $\tilde{P}$. All the coefficients in (\ref{nonres}) can be defined from comparison with experiment (here~\cite{estrecho2021,parish}). The main difference of the model (\ref{nonres}) compared to the previously developed phenomenological approaches is that it describes the dynamical feeding of the polariton system with the total density $n$, wherein the equilibration between the condensate $n_0$ and non-condensate $n^\prime$ particles is assumed to be governed by temperature. Importantly, the simplicity of Eqs.~(\ref{nonres}) allows analytical solution and does not require numerical simulation. Typical solutions $n(t)$ and $n_D(t)$ for two different pump powers are shown in Fig.~\ref{fig1} for pulsed (b) and c.w. (c) excitations.

\begin{figure}[b]
\includegraphics[width=\columnwidth]{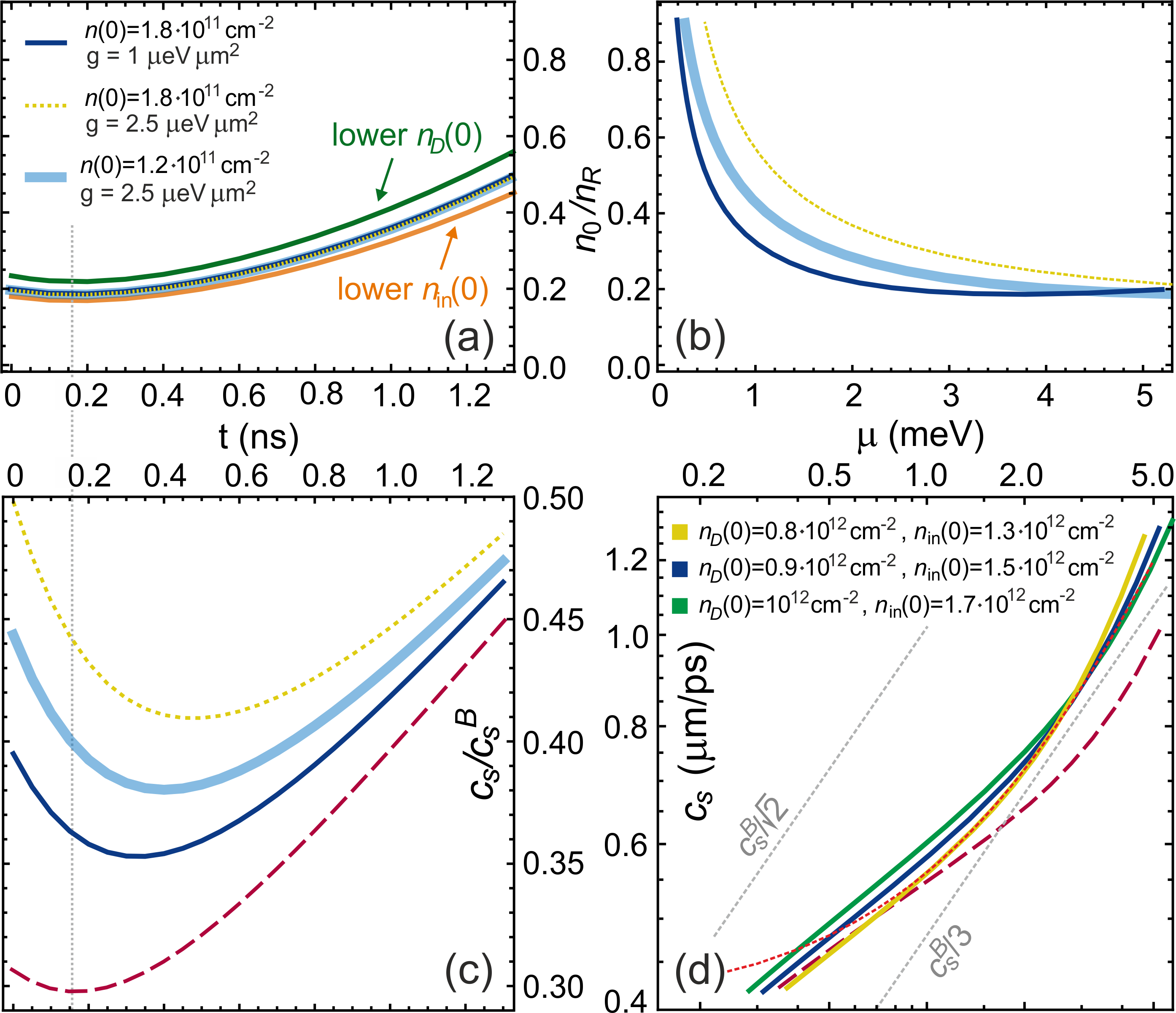}
\caption{\small {\bf Pulsed excitation.} (a) Evolution of the condensate-to-reservoir ratio $n_0/n_R$, where $n_D(t)$ and $n(t)$ are acquired from (\ref{nonres}), $n_0$ and $n^\prime$ for each $n$ are calculated within the HFB theory, and $n_R=n^\prime+n_D$. (b) $n_0/n_R$ versus the observable blueshift $\mu$ for each moment of time during the system evolution. Note that smaller values of $\mu$ correspond to later times. (c) Temporal dependence of the ratio $c_s/c_s^B$, with $c_s$ given by (\ref{cs_prb}) and the Bogoliubov value $c_s^B=\sqrt{\mu/m_{\rm LP}}$. In (a--c), different lines are plotted for two different exciton interaction strengths $g$ and initial densities $n(0)$ as indicated in (a), $n_D(0)=0.9\times10^{12}$~cm$^{-2}$, $n_{in}(0)=1.5\times10^{12}$~cm$^{-2}$.
(d) The sound velocity $c_s$ vs. $\mu$ showing the deviation from the square-root dependence (the gray dotted lines) on the blueshift. The red dotted line displays the linear fit $c_0+c_1\mu$ suggested in~\cite{estrecho2021} to match the experiment, where $c_0=0.4~\mu$m~ps$^{-1}$, $c_1=0.16~\mu$m~ps$^{-1}$meV$^{-1}$. Solid lines of different colors correspond to different initial reservoir densities (as marked) with the same $n(0)=1.8\times10^{11}$~cm$^{-2}$ and $g=1~\mu$eV~$\mu$m$^2$. The dark-red dashed lines in (c,d) show the corresponding dependencies for $c_s^0$ according to Ref.~\cite{solnyshkov}. In all panels, the decay and scattering rates are the same as in Fig.~\ref{fig1}b, $\hbar\Omega=15.8$~meV, $\Delta=1$~meV, $m_{\rm LP} = 8.6\times10^{-5}m_0$, $T=10$~K.}
\label{fig2}
\end{figure}

For the case of pulsed excitation, to be able to apply Eqs.~(\ref{cs_prb}), (\ref{mu_prb}) one has to assume that the decay of the system after the pulse arrival is slow compared to thermalisation time, so that at each moment of time the polariton subsystem may be considered in quasi-equilibrium. This is realised for high-quality samples and(or) for positive detunings $\Delta$. In this case, we set $P=\tilde{P}=0$ in (\ref{nonres}), introduce the initial values of all densities $n_{in}(0)$, $n_D(0)$, $n(0)$ and consider the scattering rate to the condensate constant: $R = R_{\rm p}$. According to the solution
\begin{equation}\label{nt}
n(t) = n(0)e^{-\gamma t} + \frac{R_{\rm p}n_{in}(0)[e^{-(R_{\rm p} + D + \gamma_{in})t}-e^{-\gamma t}]}{\gamma-\gamma_{in}-D-R_{\rm p}}
\end{equation}
shown in Fig.~\ref{fig1}b, the dynamics of the polariton density consists of an early-time faster decay and then a slower decrease at later times due to the replenishment of polaritons from the long-lived inactive reservoir. The dark-exciton reservoir population $n_D$ is decaying much slower than the polariton density $n$ [$n_D(t)$ is obtained from (\ref{nt}) by exchanging $n(0),\gamma\to n_D(0),\gamma_D$ and $R_{\rm p}\leftrightarrow D$].
At each $t$, one can directly use $n_D$ to find the dark exciton contribution to the blueshift (\ref{mu_prb}). To define the contributions of the condensate $n_0$ and the bright non-condensate excitons $n_X^\prime$, for each $n(t)$ we perform the self-consistent calculation at a given temperature (here, $T=10$~K) according to the approach developed in~\cite{grudinina}. This procedure allows us to retrieve the dependence of the condensate-to-reservoir ratio 
on time which is shown in Fig.~\ref{fig2}a. At all times the reservoir density in the system is dominant (with $n_R/n_0$ decreasing from $\approx 5$ to $\approx2$ at later times) due to the large dark population $n_D$, whereas the thermal (bright) part of the reservoir $n^\prime$ is small compared to $n_0$. Calculating the blueshift (\ref{mu_prb}) at each $t$ with the obtained densities, one can plot $n_0/n_R$ versus $\mu$ (see Fig.~\ref{fig2}b). Obtained trends are similar for all considered pump powers and interaction strengths.

Now one can track the dependence of the sound velocity in the polariton fluid both on time and on the blueshift $\mu$. In the equilibrium case at $T=0$, one expects the square-root scaling predicted by Bogoliubov~\cite{bogoliubov}: $c_s^B=\sqrt{\mu/m_{\rm LP}}$, while for the driven-dissipative case Ref.~\cite{solnyshkov} reported $c_s^0=\sqrt{gX_0^4n_0/m_{\rm LP}}$. The calculations here are performed according to (\ref{cs_prb}). In Fig.~\ref{fig2}c we plot the ratio $c_s/c_s^B$ together with $c_s^0/c_s^B$ (the dark-red dashed line). It is evident that at early times $c_s$ is rapidly decreasing from the values of the order 0.4--0.5~$c_s^B$ with a pronounced minimum, and later recovers up to values higher than those at the beginning. Notably, for $c_s^0/c_s^B$ the minimum coincides with the minimum of $n_0/n_R$ independent of the interaction strength or density (the gray dotted line in Fig.~\ref{fig2}a,c is a guide to the eye), while for $c_s$ according to (\ref{cs_prb}) the minimum shifts towards later times for higher $g$ and higher initial densities.
The dependence $c_s(\mu)$ is presented in Fig.~\ref{fig2}d for a fixed $g$. At larger blueshifts (early times) $c_s\approx c_s^B/3$ and shows a slight deviation from the square-root dependence on $\mu$ due to the renormalisation of the lower-polariton mass reported in Ref.~\cite{grudinina}. This deviation towards higher values at large $\mu$ appears regardless of the dark reservoir contribution. At later times (corresponding to $\mu<2$~meV), one sees the complete change of behavior with $c_s$ increasing from the value $c_s^B/3$ and following a linear rather than a square-root-like dependence on $\mu$. In Fig.~\ref{fig2}d, the red dotted line shows an analytical fit $c_0+c_1\mu$ suggested in Ref.~\cite{estrecho2021} to describe experimental data. The 
dependence $c_s^0(\mu)$ shows agreement with our calculation at small $\mu$ and a considerable lowering at larger values of the blueshift. This indicates that at times when the reservoir density is large, its influence 
on the sound velocity is significant.

\begin{figure}[t]
\includegraphics[width=\columnwidth]{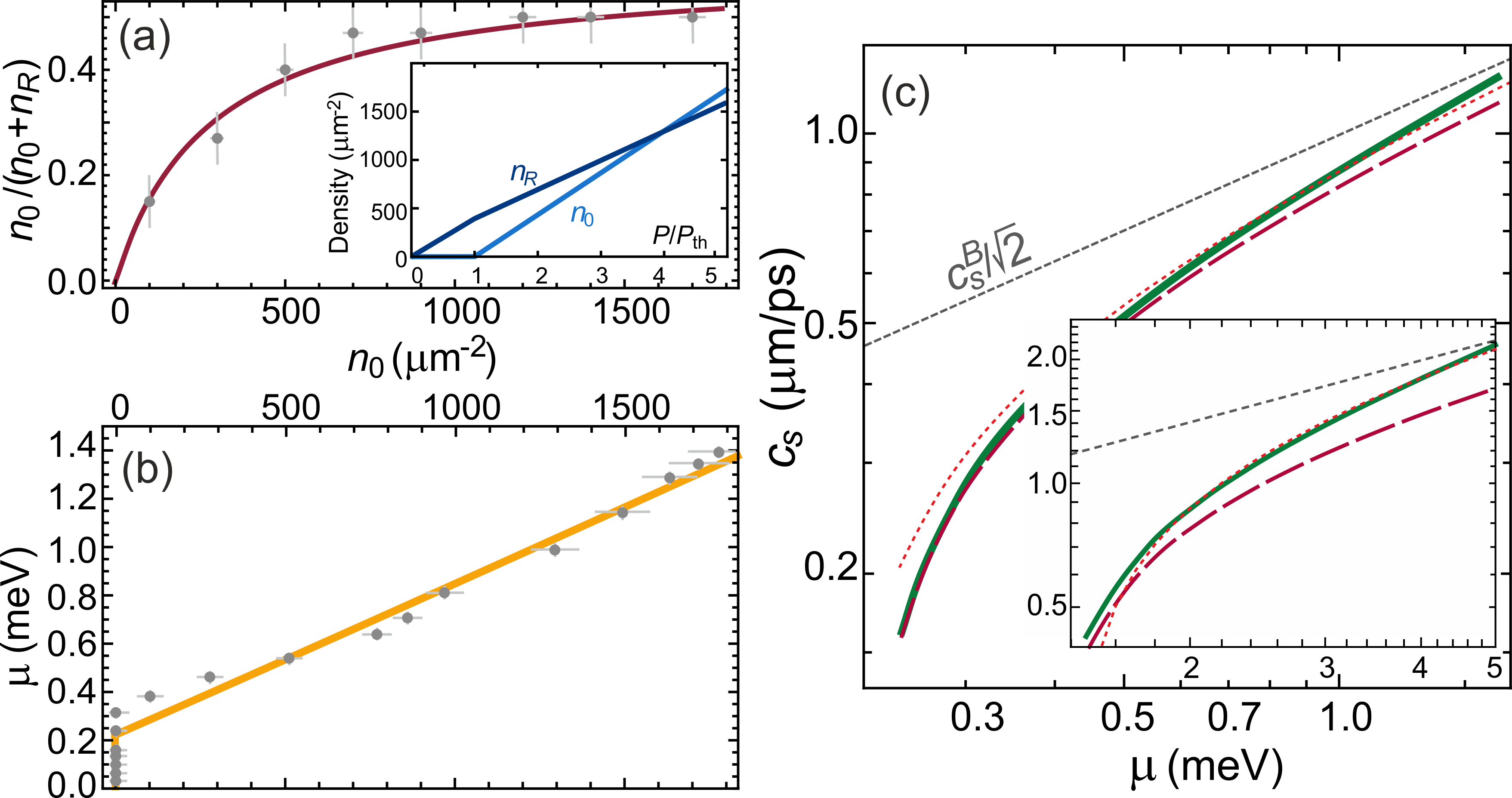}
\caption{\small {\bf Continuous-wave excitation.} (a) Condensate fraction as a function of $n_0$. The inset shows the condensate and reservoir densities vs. pump power. (b) The blueshift dependence on $n_0$. In (a,b) the gray dots with error bars show the experimental data of Ref.~\cite{parish} which were used to retrieve the pump, decay and scattering rates in Eqs.~(\ref{nonres}):
$\tilde{P}=0.25P$, $R_{\rm cw} = 0.1\times10^{-5}~\mu$m$^2/$ps, with the other parameters as in Fig.~\ref{fig1}. (c) The dependence of the sound velocity $c_s$ on the observed blueshift $\mu$ for $g=1~\mu$eV~$\mu$m$^2$. Gray dotted line shows the dependence $\propto\sqrt{\mu}$. The dark-red dashed line represents $c_s^0(\mu)$, the analytical fit is $\sqrt{0.45(\mu-0.2)/m_{\rm LP}}$ (red dotted line). The inset shows the same for larger blueshifts, by setting $g=6~\mu$eV~$\mu$m$^2$. The analytical fit is $\sqrt{0.6(\mu-1.4)/m_{\rm LP}}$.}
\label{fig3}
\end{figure}
We now turn to the case of c.w. excitation, where the system reaches the state of dynamical equilibrium at a given pump power $P$. Since the polariton density is not decaying, we assume the scattering rate between the inactive reservoir and the polaritons density-dependent $R=R(n)$ \cite{wouters2007} and consider the simplest scenario $R(n)= R_{\rm cw} n$. Comparing the steady-state solutions of (\ref{nonres}) with zero initial conditions at different values of $P$
\begin{equation}\label{cw}
n = \frac{P}{\gamma} - \frac{\gamma_{in}+D}{R_{\rm cw}},\quad n_D = \frac{\tilde{P}}{\gamma_D} + \frac{\gamma}{\gamma_D}\frac{D}{R_{\rm cw}}
\end{equation}
(above threshold) with the experimental data of Ref.~\cite{parish} (gray dots in Fig.~\ref{fig3}a,b), we arrive at the same decay and scattering rates characteristic for the sample of  Refs.~\cite{pieczarka2020,estrecho2021,parish}, which underlines the consistency of our model. For each $n$, we use the finite-temperature HFB model to calculate $n_0$ and $n^\prime$.
The dependence of the reservoir $n_R=n_D+n^\prime$ and condensate $n_0$ densities on the pump power shows a clear threshold behavior (see the inset of Fig.~\ref{fig3}a), with $P_{th} = \gamma(\gamma_{in} + D)/R_{\rm cw}$. Fig.~\ref{fig3}a and~b display the resulting dependencies of the condensate fraction $n_0/(n_0+n_R)$ and blueshift $\mu$ on $n_0$, respectively. From Fig.~\ref{fig3}a one sees that in this setting, contrary to the pulsed-excitation case, the condensate-to-reservoir ratio is much larger, saturating at $n_0\approx n_R$ at high pump powers. Substituting the densities into Eqs.~(\ref{cs_prb}),~(\ref{mu_prb}), we plot in Fig.~\ref{fig3}c the sound velocity dependence on the blueshift. Similarly to the case of pulsed excitation, $c_s$ is lowered compared to $c_s^B$ due to the presence of the reservoirs. The dependence $c_s(\mu)$ can be approximately fitted by $\sqrt{c_1\mu+c_2}$ (red dotted line in Fig.~\ref{fig3}c), showing that the sound velocity grows faster at smaller $\mu$ when the reservoir population is dominant. The dark-red dashed line shows the corresponding dependence $c_s^0(\mu)$ which, like in the previous case, goes lower than (\ref{cs_prb}) at larger blueshifts.

\begin{figure}[t]
\includegraphics[width=\columnwidth]{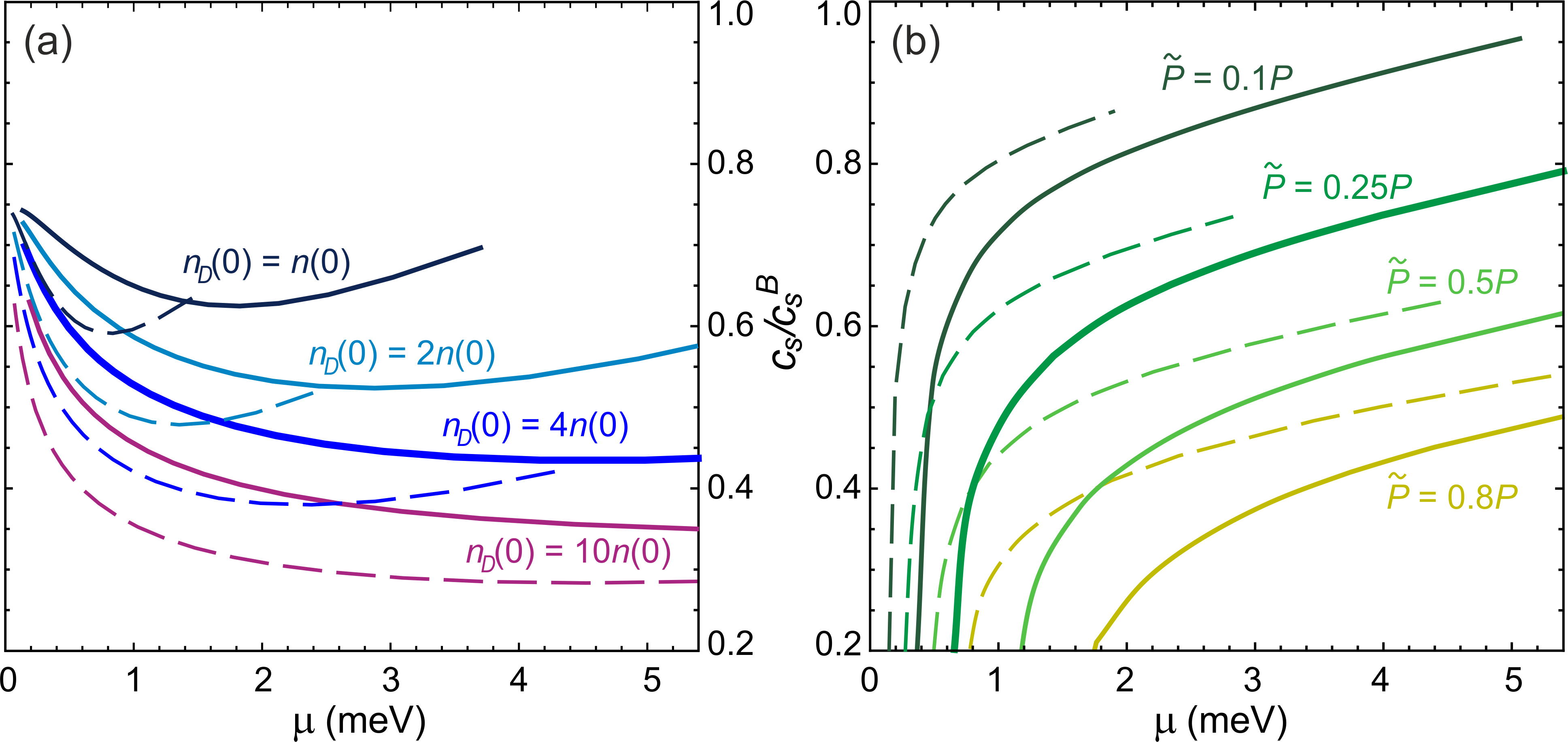}
\caption{\small Sound velocity ratio to the Bogoliubov value $c_s/c_s^B$ dependency on the observable blueshift for pulsed (a) and continuous-wave (b) non-resonant excitation schemes with different contributions from the dark exciton reservoir, as marked. The solid (dashed) lines are for $g=2.5~(1)~\mu$eV~$\mu$m$^2$.
In (a), $n(0)=1.8\times10^{11}$~cm$^{-2}$, $n_{in}(0)=1.5\times10^{12}$~cm$^{-2}$. 
All other parameters in both panels are the same as in Fig.~\ref{fig2},~\ref{fig3}. 
}
\label{fig4}
\end{figure}
Finally, we summarize in Fig.~\ref{fig4} the behaviors of the sound velocity ratio to $c_s^B$ for both types of excitation, varying the dark exciton population at a fixed pump power. One can see that compared to the case $n_D\ll n$ where $c_s$ can be slightly smaller or larger than $c_s^B$ as reported in~\cite{grudinina}, the experimentally-relevant densities of dark excitons dramatically alter the expected polariton sound velocities. As discussed above, the dependence $c_s(\mu)$ is most sensitive to the ratio between the condensate and the reservoir densities in the system. In the case of pulsed pumping scheme (Fig.~\ref{fig4}a), this ratio changes with time and the dependence $c_s(\mu)$ is non-monotonous, while staying in the range $0.3$--$0.7c_s$ for all considered $n_D$. For c.w. excitation (Fig.~\ref{fig4}b), just above threshold (i.e. at small blueshifts and small $n_0/n_R$) the sound velocity is very low, growing rapidly with $P$ up to the values $0.5$--$0.85c_s^B$ as the condensate fraction increases. Changing the interaction constant or (proportionally) all densities at the same time effectively squeezes or stretches the curves along the horizontal axis, defining the slope of the resulting dependence $c_s/c_s^B(\mu)$. Fig.~\ref{fig4} thus shows families of curves, particular cases of which were plotted in Figs.~\ref{fig2}d and~\ref{fig3}c as dependencies $c_s(\mu)$. Since the decay and scattering rates in (\ref{nonres}) can be found by fitting the observed condensate photoluminescence using the expressions (\ref{nt}) or (\ref{cw}), mapping of experimentally-measured dependence $c_s(\mu)$ in a relatively wide range of blueshifts suggests a way to define the content of dark excitons in the system at a given temperature and pump power~\cite{footnote}.

In conclusion, we developed an intuitive, analytically-solvable model based on the self-consistent HFB approach and rate equations that allows investigating the influence of dark excitons and the bright, thermally-distributed polaritonic reservoir on the sound velocity in nonresonantly-pumped polariton fluids. The only input parameters used are the decay and scattering rates that can be found from fitting the experimental data. Our model allows us to directly obtain the condensate-to-reservoir ratio, blueshift, and the sound velocity for both pulsed and c.w. excitation schemes. We show that the sound velocity is dramatically affected when the dark exciton density is large, and that it cannot be expected to follow the square-root-like scaling with the blueshift predicted by the equilibrium Bogoliubov theory. Most strikingly, since in polariton experiments the blueshift and the slope of the low-momenta elementary excitation spectrum can be measured independently, the mapping of such measurements on the dependence $c_s(\mu)$ can be used to {\it define} the density of dark excitons which is generally elusive for observations by optical means.

\begin{acknowledgements}
The authors are thankful to I.~L.~Kurbakov, E.~Estrecho, M.~Pieczarka and O.~Bleu for fruitful discussions. The work of the authors is financially supported by the Russian Foundation for Basic Research Grant No.~21--52--12038 and MEPhI Program Priority 2030.
\end{acknowledgements}


\begin{thebibliography}{99}
\bibitem{footnote}
Precisely, to be able to map the density of dark excitons in the system, one needs to fit simultaneously the two experimentally-measured dependencies: for the case of pulsed excitation, the time-dependent condensate photoluminescence (PL) providing $n_0(t)$ 
and the dependence $c_s(\mu)$, and for the case of c.w. excitation, the steady-state PL dependent on the pump power $n_0(P)$ 
or on the blueshift $n_0(\mu)$ 
together with $c_s(\mu)$, given that the sound velocity $c_s$ and the condensate blueshift $\mu$ are measured independently. Tracking the dependence $c_s(\mu)$ is crucial since it is highly sensitive to all the model parameters.
\end{thebibliography}
 \end{document}